\documentclass[prl,twocolumn]{revtex4}
\usepackage{graphicx}
\usepackage{amssymb,amsmath,bm}
\usepackage{enumerate}

\newcommand{\be}{\begin{eqnarray}}
\newcommand{\ee}{\end{eqnarray}}

\begin{document}

\title{Distributed chaos and helicity in turbulence}

\author{A. Bershadskii}

\affiliation{
ICAR, P.O. Box 31155, Jerusalem 91000, Israel
}

\begin{abstract}

The distributed chaos driven by Levich-Tsinober (helicity) integral: $I=\int \langle h({\bf x},t)~h({\bf x}+{\bf r}, t)\rangle d{\bf r}$ has been studied. It is shown that the helical distributed chaos can be considered as basis for complex turbulent flows with interplay between large-scale coherent structures and small-scale turbulence, such as Cuette-Taylor flow, wake behind cylinder and turbulent flow in the Large Plasma Device (LAPD) with inserted limiters.      In the last case appearance of the helical distributed chaos, caused by the limiters, results in improvement of radial particle confinement.   
\end{abstract}

\maketitle

\section{Introduction}

In some turbulent flows large-scale vortex structures and waves, which appear at low Reynolds numbers as a result of hydrodynamic instabilities, are still present in the flows at rather large (turbulent) values of Reynolds number (usually as coherent structures). 
Cuette-Taylor flow and wake behind cylinder are good examples of such flows \cite{my1}. At large values of Reynolds number a combination of small scale turbulence with the inherited large-scale structures provides a very interesting field for investigation. 

 At small values of Reynolds number a transition to turbulence via a chaotic scenario was discovered experimentally for the Cuette-Taylor flow: the strange attractors and characteristic exponential spectra were observed there \cite{swinney1}. But how is the chaos developing with increasing Reynolds number? The attempts to reach scaling spectrum at the laboratory experiments were unsuccessful.        
 About the same can be said on the wake behind cylinder \cite{s1}. The only difference is that at the axis of symmetry of the wake presence of the scaling spectra was observed, due to specific symmetry conditions (these spectra can be related to helicity dynamics \cite{b}). But when one steps out of the axis of the symmetry the scaling disappears. 
 
 Since the homogeneous and isotropic turbulence has its roots in the distributed chaos \cite{b1} one can look at the distributed chaos as a possibility for development of the chaos in the Cuette-Taylor and in the wake behind cylinder at the Reynolds number increase. Spectrum of the distributed chaos is a superposition of the exponentials 
$$
E(k ) \simeq \int_0^{\infty} \mathcal{P} (\kappa)~ e^{-(k/\kappa)}  d\kappa  \propto \exp-(k/k_{\beta})^{\beta}  \eqno{(1)}
$$
where $\kappa$ is wavenumber of the pulses (waves) driving the chaos and $\mathcal{P}(\kappa )$ is a distribution of the wavenumber $\kappa$. The distributed chaos is a probabilistic extension of the ordinary deterministic chaos with exponential spectrum. Therefore it is a natural candidate for the chaos development in the flows with transition to the turbulence via chaos. For the homogeneous and isotropic turbulence the distributed chaos appears without the transitional stage characterized by simple exponential spectrum. But in more complex flows there is a widening of the distribution  $\mathcal{P} (\kappa)$ (starting from a delta-function) with increase of the Reynolds number.
 
\section{Helicity conservation and distributed chaos}

 When the widening process reaches an equilibrium the parameter $\beta$ in Eq. (1) can be determined from an asymptotic theory \cite{b1}:
$$
\beta =\frac{2\alpha}{1+2\alpha}   \eqno{(2)}
$$ 
where $\alpha$ is given by scaling of the group velocity, $
\upsilon (\kappa )$, of the waves driving the chaos
$$
\upsilon (\kappa ) \sim \kappa^{\alpha}     \eqno{(3)}
$$ 
at $\kappa \rightarrow \infty $.

 In order to find the scaling exponent $\alpha$ itself the hydrodynamic invariants then can be used. For the homogeneous and isotropic turbulence the Birkhoff-Saffman invariant
$$   
I_2 = \int  \langle {\bf u} ({\bf x},t) \cdot  {\bf u} ({\bf x} + {\bf r},t) \rangle d{\bf r}  \eqno{(4)}
$$
was found to be a relevant one \cite{b1}. This invariant is associated with the momentum conservation law \cite{saf},\cite{fl},\cite{d} and, consequently, with the space translational symmetry \cite{ll2}. Substituting the the Birkhoff-Saffman integral into Eq. (3) and using the dimensional considerations we obtain $\alpha = 3/2$ \cite{b1}:
$$
\upsilon (\kappa ) \propto ~I_2^{1/2}~\kappa^{3/2} \eqno{(5)}
$$
and then, from Eq. (2): $\beta =3/4$. \\

  For the more complex flows, which are considered in present paper (see Introduction), certain process of homogenization and isotropization should take place before the flows can reach (may be locally) this state. Moreover, for the intermediate values of Reynolds number an intermediate equilibrium state, driven by another invariant, can be reached. Taking into account the inherited large-scale coherent structures in these flows one can consider the helicity analogue of the Birkhoff-Saffman integral
$$   
I = \int  \langle  h ({\bf x},t) \cdot   h ({\bf x} + {\bf r},t) \rangle d{\bf r}  \eqno{(6)}
$$
as a possible candidate. It was shown by E. Levich and A. Tsinober in Ref. \cite{lt} (see also Ref. \cite{fl}) that for helicity $h= {\bf u} \cdot (\nabla \times {\bf u})$ the integral $I$ Eq. (6) is an inviscid invariant of the Navier-Stokes equation. Substituting the Levich-Tsinober integral $I$ into the Eq. (3) and using the dimensional considerations we obtain $\alpha = 1/4$:        
$$
\upsilon (\kappa ) \propto ~I^{1/4}~\kappa^{1/4} \eqno{(7)}
$$
and then, from Eq. (2): $\beta =1/3$. 

\section{Comparison with the laboratory experiments}  

The Cuette-Taylor is a flow between concentric cylinders with a rotating inner cylinder. The Reynolds number for this flow is usually defined as $Re= \Omega a (b-a)/\nu$, where $\nu$ is the kinematic viscosity, $a$ and $b$ are the inner and outer cylinder radii and $\Omega$ is the inner cylinder angular rotation rate. 
\begin{figure}
\begin{center}
\includegraphics[width=8cm \vspace{-1cm}]{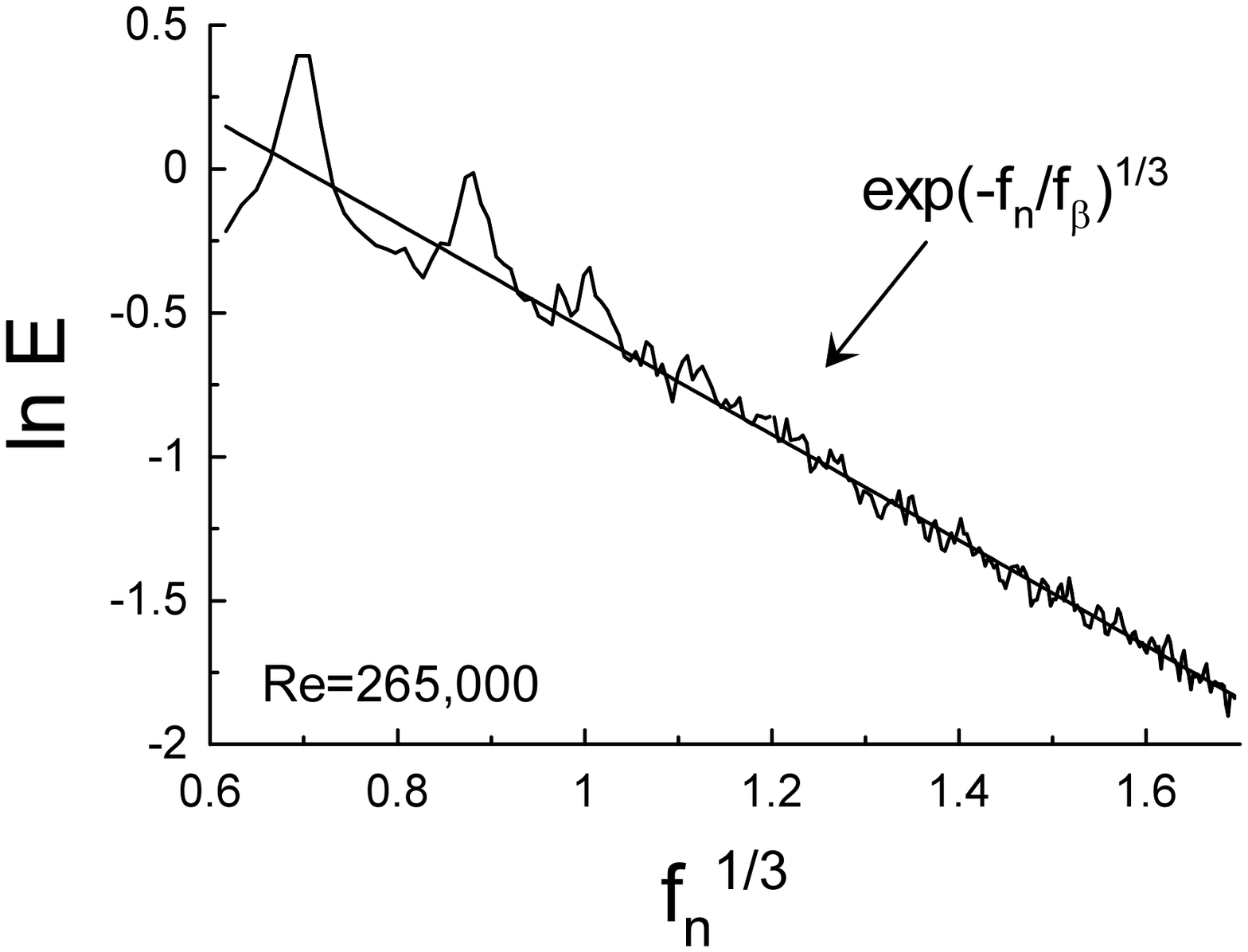}\vspace{-4cm}
\caption{\label{fig1} Logarithm of power spectrum of the azimuthal velocity at $Re=265,000$ as function of $f_n^{3/4}$, where the normalized frequency $f_n=2\pi f/\Omega$. The straight line indicates the stretched exponential decay Eq. (1) with $\beta =1/3$. }
\end{center}
\end{figure}

\begin{figure}
\begin{center}
\includegraphics[width=8cm \vspace{-1.38cm}]{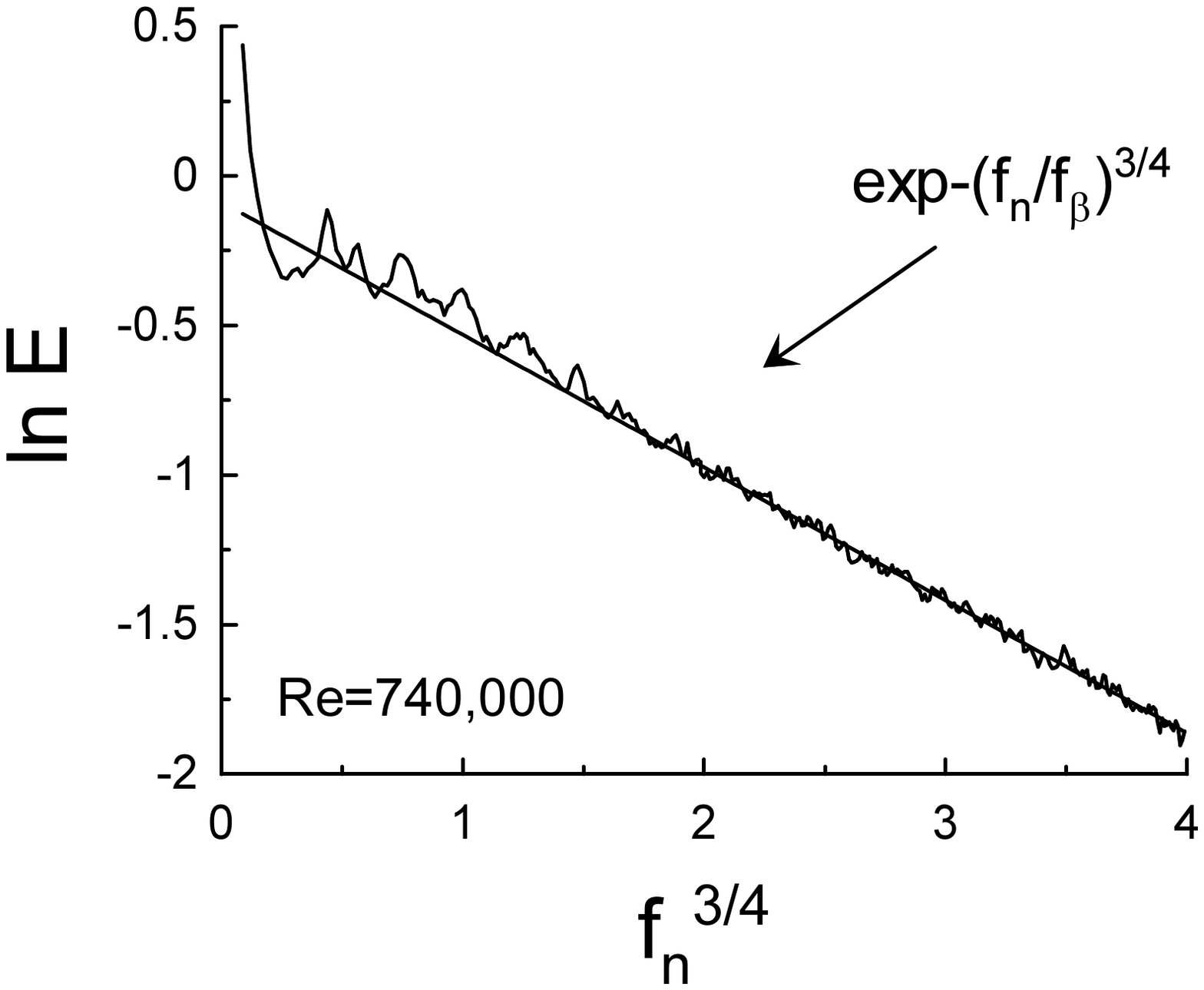}\vspace{-3.2cm}
\caption{\label{fig2} The same as in Fig. 1 but for $Re=740,000$. The straight line indicates the stretched exponential decay Eq. (1) with $\beta =3/4$.} 
\end{center}
\end{figure}

\begin{figure}
\begin{center}
\includegraphics[width=8cm \vspace{-1.2cm}]{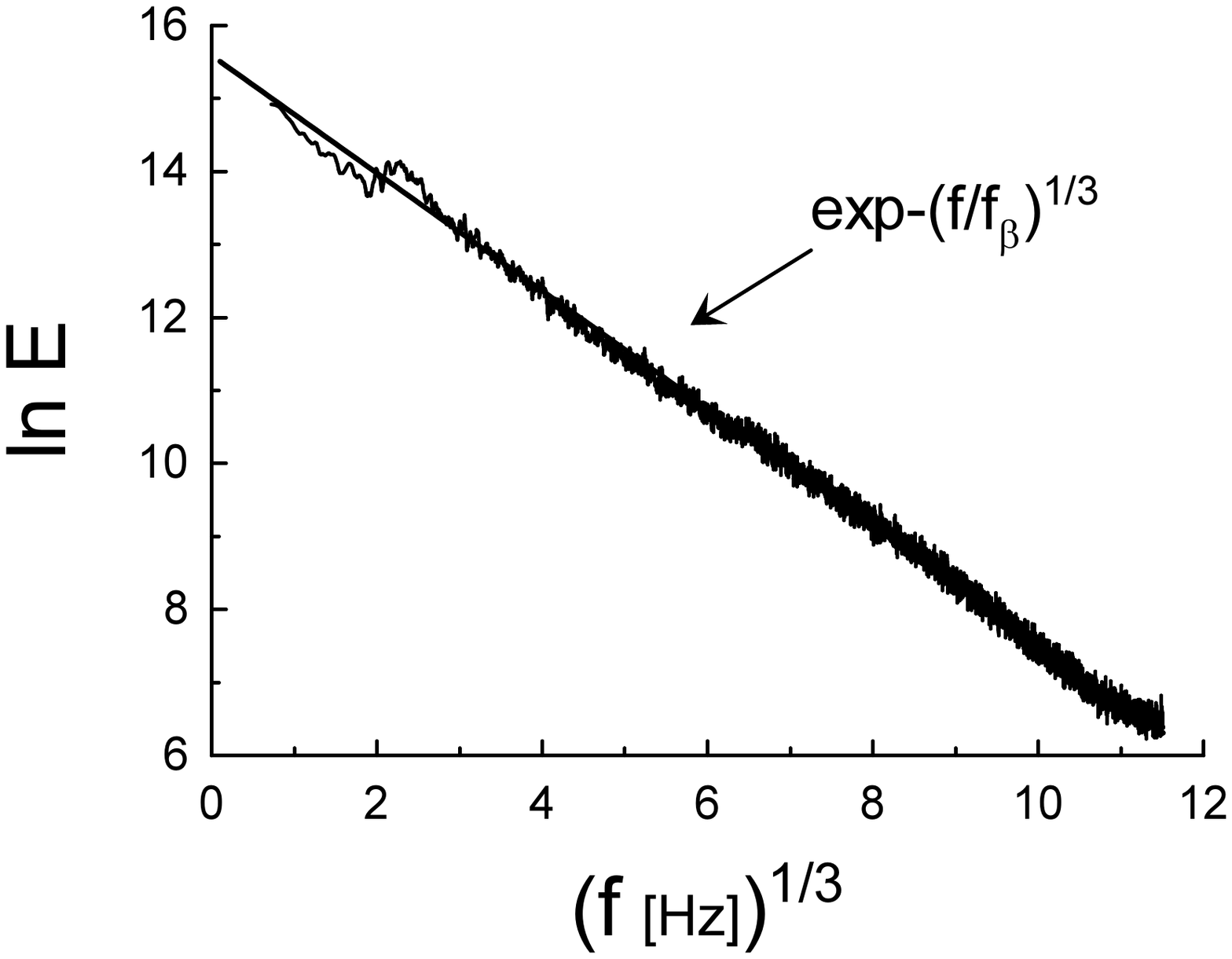}\vspace{-3.8cm}
\caption{\label{fig3} Logarithm of the power spectrum of the temperature fluctuations in the wake of a heated cylinder in a wind tunnel (at distance of three cylinder radii from the wake symmetry axis) \cite{kss}. }
\end{center}
\end{figure}
  Figures 1 and 2 show frequency power spectra of azimuthal velocity observed in a laboratory experiment with a Cuette-Taylor flow at $Re=265,000$ and $Re=740,000$ \cite{ls}. These figures represent corresponding frequency spectra ($f_n=2\pi f/\Omega$) shown in Fig. 6 of the Ref. \cite{ls}, but the scales here are chosen in order to show the stretched exponential decay Eq. (1) with $\beta =1/3$ and $\beta=3/4$ as the straight lines. The well known Taylor hypothesis \cite{my2} states that the frequency spectra, measured by a probe with a fixed space location, reflex corresponding space (wavenumber) spectra of the space structures moving (sufficiently fast) past the probe, so that: $f \propto k$ in these spectra. Measurements were made in the fluid core. The apparent peaks in the spectra correspond to the large-scale (inherited) azimuthal travelling waves, which appear at small values of $Re$. The authors of the Ref. \cite{ls} also reported "that vortex-like coherent
structures persist with increasing Re, even to $Re=10^6$". These observations (see also Ref.\cite{l}), together with the value $\beta=1/3$ (Fig. 1) support existence and domination of the helicity (Levich-Tsinober) attractor of the distributed chaos at the intermediate value of the Reynolds number $Re=265,000$. With further increase of $Re$ local \cite{my2} homogenization and isotropization of the flow results (at $Re=740,000$) in the stretched exponential spectrum with $\beta =3/4$ shown in Fig. 2 (indicating domination of the Birkhoff-Saffman attractor at this large value of $Re$).\\

  This approach can be also applied to the data acquired with a cold-wire
probe in the wake of a slightly heated circular cylinder in a wind tunnel (see Ref. \cite{kss} for
details of the experiment, Taylor length Reynolds number $Re_{\lambda} = 130$).
Temperature can be considered as a passive scalar at the applied conditions. These measurements
were temporal records taken by the probe located at distance of about three cylinder's radii from the symmetry axis of the wake. Figure 3 shows the frequency power spectrum of the temperature fluctuations from this experiment \cite{kss}. The scales are chosen in order to show correspondence of the data to the stretched exponential spectrum Eq. (1) with $\beta=1/3$. The domination of the helicity (Levich-Tsinober) attractor of the distributed chaos can be clear seen.\\

\begin{figure}
\begin{center}
\includegraphics[width=8cm \vspace{-1cm}]{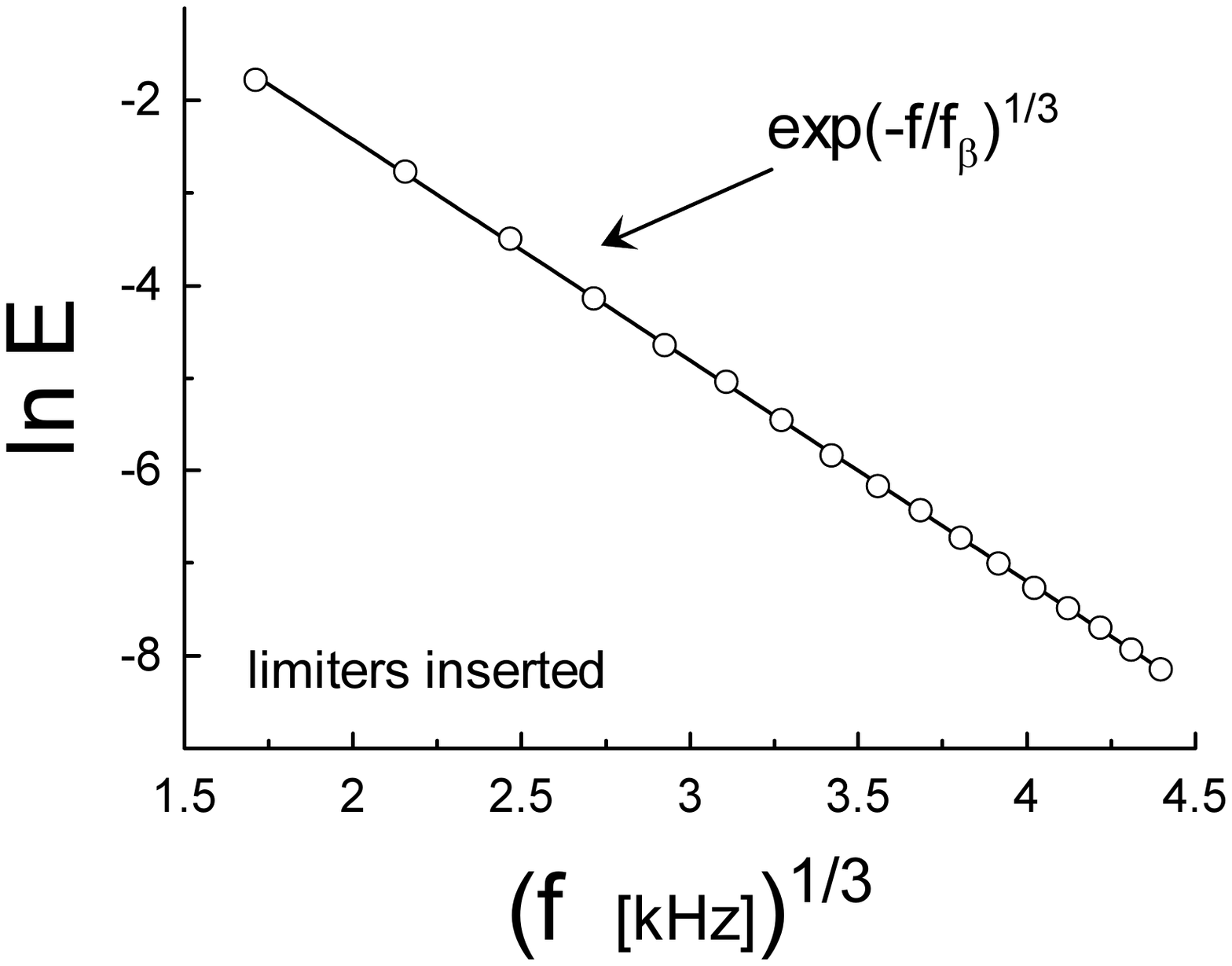}\vspace{-4cm}
\caption{\label{fig4} Logarithm of the density edge power spectrum as function of $f^{1/3}$. } 
\end{center}
\end{figure}
The Large Plasma Device produces a cylindrical magnetized plasma
column (60 cm diameter, 17 m long). In recent experiment an interplay between azimuthal flow and
turbulent transport was investigated using limiters modifying rotation of the plasma column \cite{s}. It is shown in this experiment that with the limiters inserted modification of the turbulent transport can result in improvement of radial particle confinement. The decrease in radial particle flux is caused mostly by suppression of the density fluctuation power at the edges. Figure 4 shows the high frequency part of the density power spectrum summed for the edge region of 20 to 35cm (the data were taken from Fig. 4.13b of the Ref. \cite{s}). In the scales of Fig. 4 the stretched exponential decay Eq. (1) with $\beta=1/3$ can be seen as as a straight line. The high frequency spectrum provides clear indication of the helical distributed chaos in the case of the limiters inserted. Direct analysis of corresponding time series, performed in the Ref. \cite{s}, reveals the Lorentzian pulses responsible for the exponential-like spectra \cite{b1},\cite{mm}.  \\

I thank G. S. Lewis and K. R. Sreenivasan for sharing their data.

\end{document}